\documentclass[prl,twocolumn]{revtex4}
\usepackage{amsmath}
\usepackage[dvips]{psfrag,graphicx}
\usepackage{verbatim}
\usepackage[usenames]{xcolor}
\usepackage{color}
\usepackage{subfigure}
\usepackage{hyperref}
\usepackage{float}

\hypersetup{
  colorlinks,
  citecolor=violet,
  linkcolor=black,
  urlcolor=blue
}

\begin{document}

\title{
Exact non-adiabatic holonomic transformations of spin-orbit qubits
}

\author{T. \v Cade\v z$^{1,2}$, J. H. Jef\mbox{}ferson$^3$, and A. Ram\v sak$^{1,4}$}
\affiliation{ $^{1}$ Jo\v zef Stefan Institute, Ljubljana, Slovenia}
\affiliation{ $^{2}$ Institute of Mathematics, Physics and Mechanics, Ljubljana, Slovenia}
\affiliation{ $^{3}$ Department of Physics, Lancaster University, Lancaster, LA1 4YB, United Kingdom}
\affiliation{ $^{4}$ Faculty of Mathematics and Physics, University of Ljubljana, Ljubljana, Slovenia}

\date{\today}

\begin{abstract}
An exact analytical solution is derived for the wavefunction of an electron in a one-dimensional moving quantum dot in a nanowire, in the presence of time-dependent spin-orbit coupling. For cyclic evolutions we show that the spin of the electron is rotated by an angle proportional to the area of a closed loop in the parameter space of the time-dependent quantum dot position and the amplitude of a fictitious classical oscillator driven by time-dependent spin-orbit coupling. By appropriate choice of parameters, we show that the spin may be rotated by an arbitrary angle on the Bloch sphere. Exact expressions for dynamical and geometrical phases are also derived.
\end{abstract}



\maketitle

{\it Introduction and motivation.-} The importance of a geometric phase factor in the adiabatic cyclic evolution of a non-degenerate quantum system was first discussed by Berry~\cite{Berry84} and later extended to adiabatic cyclic evolution of a degenerate quantum system, for which the acquired geometric phase is non-Abelian~\cite{Wilczek84}. Generalization to non-adiabatic cyclic evolutions was subsequently given for both non-degenerate~\cite{Aharonov87} and degenerate quantum systems~\cite{Anandan88}.

Since the original proposal of using the spin of an electron confined in a quantum dot (QD) as a qubit~\cite{Loss98}, a great deal of experimental and theoretical progress has been made on the road to realizing a quantum computer utilizing spins in QDs~\cite{Kloeffel13}.

A successful way to achieve single spin manipulation is by employing electric-dipole induced spin resonance (EDSR)~\cite{Rashba03, Rashba08, Golovach06, Kato03, Laird07, Pioro-Ladriere08, Li13, Yang13} using time-dependent electric fields, applied via gate electrodes. EDSR mediated by the spin-orbit interaction (SOI)~\cite{Golovach06} enables single spin manipulation as demonstrated in lateral QDs~\cite{Nowack07} and in QDs formed in nanowires~\cite{Nadj-Perge10, Schroer11, Nadj-Perge12}. Spin-flip times in such schemes are about 100ns in lateral QDs~\cite{Nowack07} and below 10ns in InAs nanowires~\cite{Nadj-Perge10}. Other theoretical proposals to exploit the SOI for single spin manipulations can be found in the literature~\cite{Coish06, Debald05, Flindt06, Shitade10, Ban12}.

Due to the spin-orbit interaction present in semiconductors~\cite{Dresselhaus55, Bychkov84}, the single-electron orbital states in a QD are spin-dependent and in the absence of a magnetic field the eigenstates are Kramers doublets, due to time-reversal invariance. A (spin-orbital) qubit can then be defined as the ground state Kramers doublet. 

If the QD is displaced, the SOI induces rotations of the spin-orbital qubit~\cite{Coish06}. Such rotations have been studied numerically~\cite{Bednarek08} and analytically in the adiabatic limit~\cite{San-Jose08, Golovach10, Prabhakar10}. By adiabatically moving the dot in closed loops (holonomies) general single-qubit manipulations can be achieved~\cite{San-Jose08, Golovach10} which, together with the use of the Heisenberg exchange interaction for two-qubit manipulations, enables holonomic quantum computation~\cite{Zanardi99, Sjoqvist12} with spins in QDs~\cite{Golovach10}. Experimental progress to realize this idea has been reported recently in quadruple QD systems~\cite{Thalineau12}.

In this Letter we propose the manipulation of a spin-orbital qubit in a QD via motion in only one (physical) dimension, {\em e.g.}, along quantum wire, using a time-dependent spin-orbit (Rashba) coupling with coupling parameter $\alpha(t)$, achieved electrically by changing the potential on a gate electrode~\cite{Nitta97}. This contrasts with previous proposals in which the parameter space is two-dimension position space~\cite{Coish06, Bednarek08, San-Jose08, Golovach10, Prabhakar10}. Recently a six-fold tuning of $\alpha$ was demonstrated within 1~V of gate bias in an InAs nanowire~\cite{Liang12}. 

To demonstrate single qubit manipulation, we give an exact analytical solution for the wavefunction of an electron in a one-dimensional moving quantum dot, modelled by a time-dependent confining harmonic potential in the presence of time-dependent spin-orbit coupling. This solution is itself interesting and adds to a limited number of exactly solvable time-dependent problems among which are general time-dependent harmonic oscillators~\cite{Cordero-Soto08, Harari11}, tunnel-coupled spin qubits driven by ac fields~\cite{Gomez-Leon12} and time-dependent two-level systems~\cite{Barnes12, Barnes13}.

{\it Model and exact solution.-} We consider the Hamiltonian of a single electron in a one-dimensional system
\begin{eqnarray}\label{Ht}
H(t) = \frac{p^2}{2  m^*} I + \frac{m^*  \omega^2}{2}  \bigl[{x} - \xi(t) \bigr]^2 I + \alpha(t)  \,p \,{\bf n} \cdot {\boldsymbol \sigma},
\end{eqnarray}
where $m^*$ is the electron effective-mass and $\omega$ is the frequency of a harmonic trap (moving QD). The momentum and position operators are $p$ and $x$, respectively. The dot is translated with time-dependent position defined by the harmonic potential minimum at $\xi(t)$. The spin rotation axis due to the SOI $\alpha(t)$ is denoted by a unit vector ${\bf n}$, which depends on the crystal structure of the quasi one-dimensional material used and the direction of the applied electric field~\cite{Nadj-Perge12}. Pauli matrices and identity in the spin space are ${\boldsymbol \sigma}$ and $I$, respectively.

Before presenting the solution, we give a simple example of the manipulation we have in mind. Consider a spin-orbital qubit in a nanowire QD with a constant Rashba SOI, $\alpha=\alpha_1$ and translation of the QD by some distance $\xi_0$. For adiabatic driving, this movement induces a rotation of the qubit on the Bloch sphere, which is proportional to the product $\xi_0 \alpha_1$~\cite{Coish06}. For example, a spin-flip can be realized if the distance travelled is $\pi m^*  \alpha/2$. This can also be achieved by non-adiabatic movement of the QD, enabling spin-flips with frequency of the order of the QD level spacing~\cite{Cadez13}. After translation, with the dot at a fixed position, the coupling is then changed from $\alpha_1$ to $\alpha_2$ with corresponding evolution of the Kramer's doublets. As we show below, the evolution of the SOI can be tuned analogously to that of QD displacement and this evolution can also be non-adiabatic. By displacing the dot back to the starting position, while keeping $\alpha_2$ fixed and finally driving $\alpha$ back to its initial value $\alpha_1$, a unitary transformation is applied to the original manifold. This transformation depends only on the area of the loop in the parameter space of both drivings. 

The exact solution of the time-dependent Schr\" odinger equation for $H(t)$ is obtained via a unitary transformation ${\cal{U}}^{\dagger}(t)$, chosen so that $H_0 = {\cal{U}}(t)  H(t)  {\cal{U}}^{\dagger}(t) - i  {\cal{U}}(t)  \dot{{\cal{U}}}^{\dagger}(t) = p^2/(2  m^*) + m^*  \omega^2  x^2/2$, {\it i.e.}, an oscillator at the origin without SOI. The unitary transformation is a combination of two transformations
\begin{eqnarray}\label{U^+}
{\cal{U}}^{\dagger}(t) = {\cal{A}}_{\alpha}  {\cal{X}}_{\xi},
\end{eqnarray}
where ${\cal{X}}_{\xi}$ is the transformation into a frame moving with the QD~\cite{Kerner58}
\begin{eqnarray}\label{V}
{\cal{X}}_{\xi} = e^{- i  \phi_{\xi}(t)  I}  e^{i  m^*  [x-x_c(t)]  \dot{x}_c(t)  I}  e^{-i  x_c(t)  p  I},
\end{eqnarray}
where the phase factor $\phi_{\xi}(t) = - \int_0^t   L_{\xi}(\tau)d\tau$ is the action integral, with $L_{\xi}(t) =m^*  \dot{x}_c(t)^2/2 - m^*  \omega^2  [x_c(t) - \xi(t)]^2/2$ the Lagrange function of a driven oscillator and $x_c(t)$ is the response to the driving $\xi(t)$, {\it i.e.}, the solution to the equation of motion of a classical  driven oscillator  
\begin{eqnarray}
\ddot{x}_c(t) + \omega^2  x_c(t)  = \omega^2  \xi(t). \label{x_c} 
\end{eqnarray}
For constant $\alpha$ the transformation is given by ${\cal{A}}_{\alpha} = e^{- i \, m^*  x  \alpha \, {\bf n} \cdot {\boldsymbol \sigma}}$ as shown in Refs.~\cite{Khaetskii00, Cadez13}. For time-dependent $\alpha(t)$ this must be generalised to 
\begin{eqnarray}
{\cal{A}}_{\alpha} &=& e^{- i  [(\phi_{\alpha}(t)+m^*  \dot{a}_c(t)  a_c(t)/\omega^2)  I + \phi(t)  {\bf n} \cdot {\boldsymbol \sigma}]}  \times \nonumber \\
&& e^{- i  \dot{a}_c(t)  p  {\bf n} \cdot {\boldsymbol \sigma}/\omega^2}  e^{- i  m^*  x  a_c(t)  {\bf n} \cdot {\boldsymbol \sigma}},
\end{eqnarray}
with another action integral phase factor $\phi_{\alpha}(t) = - \int_0^t   L_{\alpha}(\tau)d\tau$, where $L_{\alpha}(t) = m^*  \dot{a}_c(t)^2/(2  \omega^2) - m^*  a_c(t)^2/2 + m^*  a_c(t)  \alpha(t)$ is the Lagrange function of another driven oscillator, satisfying
\begin{eqnarray}
\ddot{a}_c(t) + \omega^2  a_c(t) = \omega^2  \alpha(t) \label{a_c}
\end{eqnarray}
and a phase factor $\phi(t) = - m^*  \int_0^t  \dot{a}_c(\tau)  \xi(\tau)d\tau$. By analogy with the transformation ${\cal{X}}_{\xi}$, we may regard this generalised ${\cal{A}}_{\alpha}$ as transforming into the "moving frame" of the spin-orbit coupling, whilst also performing a momentum-dependent spin rotation. The phase $\phi(t)$ is a crucial term that rotates the spin and we will focus on it later. Note also the equivalence of classical equations of motion for driven oscillators Eq.~\eqref{x_c} and Eq.~\eqref{a_c}, thus the analysis of the classical driving of position~\cite{Cadez13} can also be applied to the driving of the SOI. 

The solution of the original Hamiltonian Eq.~\eqref{Ht}, $| \Psi(t) \rangle$, is obtained directly via the unitary transformation Eq.~\eqref{U^+}, {\it i.e.}, $| \Psi (t) \rangle = {\cal{U}}^{\dagger}(t)  | \psi (t) \rangle$, where $| \psi (t) \rangle$ is a solution of the transformed Hamiltonian, $H_0$. Thus we have
\begin{eqnarray}\label{wavefunction2}
| \Psi (t) \rangle &=& U(t)  | \Psi (0) \rangle,
\end{eqnarray}
where $U(t) = {\cal{U}}^{\dagger}(t)  e^{- i  H_0 t}  {\cal{U}}(0)$ is the time evolution operator. For the cases when the initial state is an eigenfunction of $H(0)$, {\it i.e.}, $| \Psi_{m s}(0) \rangle = {\cal{U}}^{\dagger}(0)  | \psi_m \rangle  | \chi_s \rangle$, where $| \psi_m \rangle$ is the $m$-th eigenfunction of the undriven harmonic oscillator $H_0$ and $| \chi_s \rangle$ is a spinor with spin $s$, the time evolved state simplifies to
\begin{eqnarray}\label{wavefunction3}
| \Psi_{m s}(t) \rangle & = & e^{-i  \omega_m  t}  {\cal{U}}^{\dagger}(t)  | \psi_m \rangle  | \chi_s \rangle,
\end{eqnarray}
where $\omega_m = (m + \frac{1}{2})  \omega$.


Henceforth we shall only consider systems with cyclic evolutions, {\it i.e.}, cases where the Hamiltonian Eq.~\eqref{Ht} after time $T$ returns to its initial form, $H(T) = H(0)$ with the state spanning the same Kramers doublet subspace defined by $m$, allowing arbitrary superpositions of the two Kramer's states. To ensure periodic behaviour the driving parameters $\xi(t)$ and $\alpha(t)$ are chosen so that $\xi(t+T)=\xi(t)$, $\alpha(t+T)=\alpha(t)$ and via the classical oscillator equations of motion Eq.~\eqref{x_c} and Eq.~\eqref{a_c}, also $x_c(t+T)=x_c(t)$ and $a_c(t+T)=a_c(t)$. This can be achieved using specific drivings in both adiabatic and non-adiabatic regimes~\cite{Cadez13}. The final state after cyclic evolution is given by Eq.~\eqref{wavefunction3} for which, at $t=T$, the time evolution operator reduces to the simple non-Abelian U(2) transformation, $U(T)=e^{ i  \Phi_T}$,
\begin{equation}
\Phi_T= [ -\omega_m  T  + \int_0^T \!\!\!L(\tau) d\tau  ]  I - \phi_T {\bf n} \cdot {\boldsymbol \sigma},\label{totalna}
\end{equation}
where $L(\tau)=L_{\xi}(\tau)+L_{\alpha}(\tau)$ is the Lagrange function of a classical two-dimensional oscillator, and the angle of spin rotation around ${\bf n}$, $2\phi_T =$ $2\phi(T)$, is given by
\begin{eqnarray}\label{phase}
\phi_T &=& m^*  \oint_{{\cal C}_1}  a_c[\xi]  {d}\xi= m^*  \! \oint_{{\cal C}_2}  \!\alpha[x_c]  {d}x_c,
\end{eqnarray}
where $a_c[\xi]$ is the response $a_c(t)$ expressed as a function of the driving $\xi(t)$ and the contour ${\cal C}_{1}$ is the path in the parametric space $[\xi(t), a_c(t)]$. Similarly,  $\alpha[x_c]$ is the driving $\alpha(t)$ expressed as a function of the response $x_c(t)$ and ${\cal C}_{2}$ is the path in the parametric space $[x_c(t),\alpha(t)]$. 

In the adiabatic limit, $x_c(t) \!\to\! \xi(t)$  and $a_c(t)\! \to\! \alpha(t)$. The total phase, Eq.~\eqref{totalna} may then be decomposed into dynamical and geometric Wilczek-Zee parts~\cite{Wilczek84}, $\Phi_T=\Phi_{\rm dyn}+\Phi_{\rm geom}$, with
\begin{eqnarray}\label{dyngeom}
\Phi_{\rm dyn} &=&- [\omega_m T - \frac{1}{2}m^*\!\!\int_0^T \!\!\alpha^2(\tau) {d} \tau] I,\\ 
\Phi_{\rm geom} &=&  -\phi_{\rm ad}  {\bf n} \cdot {\boldsymbol \sigma}, \quad \phi_{\rm ad}=m^* \! \oint_{{\cal C}_{\rm ad}} \!\! \alpha[\xi]  {d}\xi.
\end{eqnarray}
Note that $\phi_{\rm ad}$ is expressed solely in terms of the driving functions and the contour ${\cal C}_{\rm ad}$ corresponds to the path in the parametric space $[\xi(t),\alpha(t)]$, similar to the case of Berry phase for a non-degenerate state \cite{Berry84}.

\begin{figure}[htb]
\centering
\includegraphics[width=0.3\textwidth]{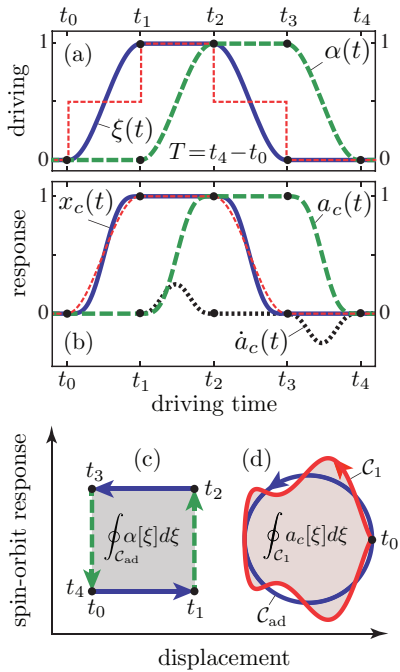}
\caption{(a) Displacement $\xi(t)$ and spin-orbit coupling $\alpha(t)$, in scaled units  as (sinusoidal) functions of driving time and (b)  scaled responses, $x_c(t)$ , $a_c(t)$ and  $\dot{a}_c(t)$  with $T=12\pi/\omega$. 
Thin dashed lines indicate discontinuous changes in displacement with $T=4\pi/\omega$ and the corresponding smooth response. 
Contours $[\xi(t),a_c(t)]$ for square and circular loops are shown in (c) and (d).  Note that the acquired phase in (c) is always unity in scaled units - the adiabatic result, as indicated. By contrast the harmonically driven case in (d) is a circle only in the adiabatic limit, with oscillatory contour otherwise. } \label{Fig:1}
\end{figure}
{\it Examples of holonomic spin manipulations.-}
The exact expression for the angle of spin rotation, Eq.~\eqref{phase}, is the central result of this paper from which pseudo-spin transformations within the Kramer's doublet may be implemented via non-adiabatic evolution. The phase is proportional to the area in the parametric space of the displacement driving and the spin-orbit response (or equivalently the spin-orbit driving and displacement response). Although the periodic driving functions are arbitrary, a particularly convenient driving scheme is to change $\xi(t)$ and $\alpha(t)$ sequentially, returning to the displaced Kramers doublet space at intermediate times, between which the evolution may be manifestly non-adiabatic with spin rotations. This is illustrated in Fig.~\ref{Fig:1}(a and b) in which at times $t_1, t_2, t_3$ and $t_4$ the response functions are equal to the driving functions, with zero time derivatives, ensuring that at this times we return to the displaced Kramers doublet space.

From these time variations of drivings and responses we see that the parametric plot of $[ \xi(t),a_c(t)]$  is a square (in scaled units), as shown in Fig.~\ref{Fig:1}(c). Comparing drivings and responses, we see that although we are in the non-adiabatic regime, from Eq.~\eqref{phase} the acquired non-Abelian phase is calculated to be $\phi_T= - m^*  \xi_0  \alpha_0$, the adiabatic result. It follows that $\phi_T$ is independent of the specific choice of drivings, provided there are no residual oscillations at the vertices. The only difference between the various cases is the cycle time, $T$, which is a minimum for the highly non-adiabatic case of instantaneous switching [thin dashed lines in Fig.~\ref{Fig:1}(a),(b)]~\cite{Cadez13}, with similar behaviour for $\alpha(t)$. Explicit calculation gives $T = 4\pi/\omega$.  An estimate for InSb gives $\phi_T \sim 1$, using parameters $m^* = 0.015 m_e$, $\xi_0 = 200$~nm and $\alpha_0 = 50$~nm$/$ps~\cite{Nadj-Perge12}. Within the expected range of allowed parameters, we see that arbitrary rotations about a fixed axis ${\bf n}$ are feasible. This rotation axis itself may be changed using additional side gates, thus opening the possibility of arbitrary rotation on the Bloch sphere.

The independence of acquired phase to the switching profile of $\xi(t)$ and $\alpha(t)$, and its equivalence to the adiabatic result, is a consequence of sequentially switching these driving terms and is not generally true.  An example is shown in Fig.~\ref{Fig:1}(d) for which $\xi(t) = \xi_0 \cos( \omega  t/n) $ and $\alpha(t) = \alpha_0 \sin( \omega  t/n)$, with integer $n \ge 2$. In the adiabatic limit ($n\to\infty$), the parametric plot is a circle when spin-orbit and displacement parameters are scaled with $\alpha_0$ and $\xi_0$. However, unlike the previous example, the trajectories and the acquired phase deviate from the adiabatic limit, as can be seen by the red contour line ${\cal C}_1$ corresponding to $n=4$~\cite{SM}. 

{\it Relation to geometric phases.-}  Anandan has extended the definitions of dynamical and geometric phases to cases of non-adiabatic cyclic evolution of degenerate systems~\cite{Anandan88}. Using our exact solution, we may derive explicit expressions for these contributions to the total phase.

We focus on cases where at $t=0$ our system is in a degenerate Kramers doublet state, Eq.~\eqref{wavefunction3}. This state undergoes cyclic evolution in time $T$ where it is again in the same Kramers doublet space but with a non-Abelian spin transformation with total phase given by Eq.~\eqref{totalna}. Note however that, due to non-adiabatic evolution, the state at intermediate times will generally be a superposition of other Kramers doublets when expressed in the instantaneous eigenbasis set at time $t$.

To divide the phase into dynamical and geometric parts we first choose a gauge transformation to another orthonormal basis~\cite{Anandan88, GeometricPhase}
\begin{eqnarray}\label{tildepsi}
| \widetilde{\Psi}_{m s} (t) \rangle &=& e^{-i \Phi(t) } | \Psi_{m s}(t)
\rangle,\\
\Phi(t)&=& [ -\omega_m t + \int_0^t \!\!\!L(\tau) d\tau ] I - \phi(t) {\bf
n} \cdot {\boldsymbol \sigma},\nonumber
\end{eqnarray}
spanning the same subspace but with periodic basis states, $| \widetilde{\Psi}_{m s} (T) \rangle = | \widetilde{\Psi}_{m s} (0) \rangle$, since $\Phi(T)=\Phi_T$, $\Phi(0)=0$. The time evolution operator, Eq.~\eqref{wavefunction2}, after one cycle can then be expressed as~\cite{Anandan88}
\begin{eqnarray}\label{U}
U(T) =e^{i \Phi_T } = {\cal{T}} e^{i \int_{0}^T [A(\tau) - K(\tau)] {d} \tau},
\end{eqnarray}
where ${\cal{T}}$ denotes the time ordering operator. Hermitian matrices $A_{s s'} = i \langle \widetilde{\Psi}_{m s} (\tau) | {d}/{d}\tau | \widetilde{\Psi}_{m s'} (\tau) \rangle$ and $K_{ s s'} = \langle \widetilde{\Psi}_{m s} (\tau) | H(\tau) | \widetilde{\Psi}_{m s'} (\tau) \rangle$ can be, by the virtue of exact functions Eq.~\eqref{wavefunction3}, expressed analytically in terms of driving and response functions. The resulting $A$ and $K$ commute allowing the total phase to be split into $\Phi_T=\Phi_{\rm dyn}+ \Phi_{\rm geom}$, where $\Phi_{\rm dyn}=-\int_{0}^T \! K(\tau){d} \tau$ and $\Phi_{\rm geom}=\int_{0}^T \! A(\tau){d} \tau$,
\begin{eqnarray}
\Phi_{\rm dyn}&=& -\int_0^T \!\!E(\tau) d\tau I-2( \phi_T-\phi_c) {\bf n} \cdot {\boldsymbol \sigma}, \label{dyn} \\
\Phi_{\rm geom}&=&\phi_{\rm a} I +( \phi_T-2\phi_c) {\bf n} \cdot {\boldsymbol \sigma},\label{geom}
\end{eqnarray}
and $E -\omega_m =m^*(\dot{x}_c^2 + \dot{a}_c^2/\omega^2)/2+ m^*\omega^2(x_c - \xi)^2/2+m^*a_c^2/2 - m^* a_c \alpha$ is instantaneous energy of driven classical oscillators, and
\begin{eqnarray}
\phi_c&=& m^* \!\oint_{{\cal C}_3} \!a_c[x_c] {d}x_c,\label{fic}\\
\phi_{\rm a}& = & m^* ( \oint_{{\cal C}_4} \!\dot{x}_c[x_c] {d}x_c + \oint_{{\cal C}_5}\!\dot{a}_c[a_c] {d}a_c /\omega^{2} ). \label{fiaa}
\end{eqnarray}
The geometric Anandan phase Eq.~\eqref{geom} is dependent on contours ${\cal C}_{1..5}$, which correspond to the trajectories of time-evolved states of the two Kramers states at $t=0$. In the adiabatic limit $\Phi_T$ reduces to the Wilczek-Zee phase for which $E(t)$ becomes the time-dependent eigenenergy of $H(t)$ and ${\cal C}_{1,2,3}\to {\cal C}_{\rm ad}$, $\phi_T\to\phi_c\to\phi_{\rm ad}$ , $\phi_a\to 0$. Note that the dynamical phase in the adiabatic limit is just the (diagonal) time-integrated energy whilst the geometric phase embodies the spin rotation, as expected. However, this is not generally the case in the non-adiabatic regime for which the spin-rotation is shared between geometrical and dynamical parts~\cite{SM}. From Eq.~\eqref{dyn} we see that when $\phi_T = \phi_c$ the spin rotation is purely geometric. In contrast from Eq.~\eqref{geom}, when $\phi_T = 2 \phi_c$, the spin rotation is purely dynamic. 

Finally, let us relate our results corresponding to the cyclic evolution of a degenerate system to results valid for non-degenerate systems. Degeneracy of the Kramers states Eq.~\eqref{wavefunction3} can be lifted, {\it e.g.}, by an external magnetic field which breaks the time reversal symmetry.  We consider the case with the magnetic field along the direction of the effective field induced by the moving QD due to SOI, {\it i.e.}, with the Zeeman term $H_z = - g \, \mu_B \, B \, {\bf n} \cdot {\boldsymbol \sigma}$. 
For more general cases the exact solution may still give new insight by treating the magnetic field component perpendicular to ${\bf n}$ as a perturbation.
The solutions to a driven Hamiltonian $H(t) + H_z$ are in this case also given by  Eq.~\eqref{wavefunction3}, but with states $|\chi_s \rangle$ being spinor eigenstates in a magnetic field, with eigenenergy $\omega_{m s} = (m + \frac{1}{2}) \, \omega \mp g \, \mu_B \, B$,  for $s=\pm \frac{1}{2}$, respectively. In this case the Aharonov-Anandan geometric phase~\cite{Aharonov87}  can also be expressed exactly, $\beta_s = \int_0^{T} \, \langle \widetilde{\Psi}_{m s} | {{d}}/{{d}}\tau | \widetilde{\Psi}_{m s} \rangle \, {\mathrm{d}}\tau=\phi_{\rm a} \pm (\phi_T-2\phi_c)$, which in the adiabatic limit reduces to the ordinary Berry phase $\beta_s \to \mp \phi_{\rm ad}$.

{\it Discussion and conclusion.-} Geometric phase has recently been measured in a driven harmonic oscillator, implemented as one of the electromagnetic modes of a transmission line resonator using a superconducting qubit as a nonlinear probe of the phase~\cite{Pechal12}. With respect to this experiment the driven harmonic oscillator considered here, Eq.~\eqref{Ht}, has an additional internal degree of freedom (spin) which the driving of the momentum couples to. This suggests the possibility of including an additional degree of freedom also in experiment, {\it e.g.}, the polarization of a photon, together with its coupling, to observe the non-Abelian phases.

Compared with other proposals, such as EDSR or inverse engineering~\cite{Ban12}, in our scheme the spin control is all-electrical without magnetic field, thus qubit errors  from fringing magnetic fields are no longer an issue. Furthermore, our general exact solution allows extensive exploration and optimisation of the model, including the non-adiabatic regime, in contrast to EDSR where SOI~\cite{Rashba03, Golovach06} or Zeeman terms~\cite{Li13} are treated as perturbations, restricting applicability.

Possible effects of the environment on the pseudo-spin state of an electron in a moving QD are decoherence and relaxation due to fluctuating electric fields, caused by the piezoelectric phonons and conduction electrons in the circuit~\cite{San-Jose08, San-Jose06}, due to hyperfine interaction with the nuclei~\cite{Echeverria-Arrondo13} or ionized dopant nuclei in a hetero-structure~\cite{Huang13}. In the last case the longitudinal and transverse rates are at the lowest order in SOI proportional to the speed of the moving QD (see Eq. (28) in Ref.~\cite{Huang13}) which applies in our case for linear ramp driving in the adiabatic limit~\cite{Cadez13}. The spin relaxation of free and QD-localized electrons with spin-orbit coupling disorder has also been studied~\cite{Glazov10} though not in the moving QDs. An important consideration in the practical implementation of this scheme is the affect of random fluctuations in both the time-dependent SOI and the QD motion. Although a detailed investigation of this is beyond the scope of the present Letter, we point out that our method and exact solution applies to arbitrary drivings of both the QD motion and time-dependent SOI and could therefore be used as a starting point for a detailed study of noise properties, which would be welcome. This would extend the already promising results for static disorder of adiabatic QD motion~\cite{Huang13} and spin-orbit coupling disorder~\cite{Glazov10}.

To conclude, we have presented a formalism for the analysis of holonomic spin-orbit qubit manipulations, where the non-Abelian U(2) phase acquired during one cycle is exactly given by the contour integral in the space of time-dependent QD position and Rashba interaction response. The time evolution operator $U(t)$ and hence also the wavefunction, is completely determined by the driving parameters $\xi, \alpha$ and their responses $x_c, a_c$. Analytical expressions derived allow a detailed analysis of different types of driving, with potential application to the design and optimisation of high-speed qubit gates. Explicit expressions for dynamic and geometric phases enable the off-diagonal (spin-rotation) part to be arbitrary, shared between them.

T. \v C. and A. R. thank Toma\v z Rejec and Vasja Susi\v c for fruitful discussions and acknowledge the support by Slovenian ARRS grant P1-0044. Support from the EU Marie Curie Network NanoCTM is also acknowledged.

\bibliographystyle{apsrev4-1}
\bibliography{/Users/cadez/work/bibliography}

%
%
\pagebreak

\begin{widetext}
\begin{center}
{{\bf Exact non-adiabatic holonomic  transformations of spin-orbit qubits\\ \ \\}}
{{\bf SUPPLEMENTAL MATERIAL \\ \ \\}}
{{T. \v Cade\v z, J. H. Jef\mbox{}ferson, and A. Ram\v sak}}
\end{center}
\end{widetext}

\setcounter{figure}{0}
\renewcommand{\figurename}{Fig.}
\renewcommand{\thefigure}{S\arabic{figure}}

Here we give two examples of periodic motion for which the drivings $\xi(t)$ and $\alpha(t)$ vary simultaneously and for which the state returns to the initial Kramers subspace after a complete period, $T$. The first example (circular driving) demonstrates the transition from highly non-adiabatic to adiabatic driving and shows clearly the effect of non-adiabatic motion on the various parametric contours. These are also seen in the second example (broken ellipsoidal driving) which can give rather exotic trajectories and for which we also show how an arbitrary division between dynamic and geometric phases can be made. This contrasts with the scenario discussed in the text for which $\xi(t)$ and $\alpha(t)$ are varied sequentially returning to the displaced Kramers subspace at intermediate times, which always gives square contours for the parametric plots, the adiabatic result.

\section{Driving and response}

The exact time evolution operator $U(t)$ is completely determined, firstly by the quantum dot time-dependent displacement coordinate $\xi(t)$ and time-dependent Rashba coupling $\alpha(t)$ and secondly by the response to these two driving functions. Displacement response $x_c(t)$ and spin-orbit response $a_c(t)$ are related to drivings by an uncoupled set of harmonic equations of motion,
\begin{eqnarray}
\ddot{x}_c(t) &+& \omega^2  x_c(t)  = \omega^2  \xi(t), \label{x_c1}\nonumber \\
\ddot{a}_c(t) &+& \omega^2  a_c(t) = \omega^2  \alpha(t). \label{a_c1}\nonumber
\end{eqnarray}
In the present case we seek periodic response to periodic driving. Of a particular interest are the solutions where the initial (and final) values of driving and response coincide. For example, when $x_c(0)=0$ and $\dot x_c(0)=0$, the solution is given by  
\begin{equation}
x_c(t)=\omega\int_0^t \sin[\omega(t-\tau)]  \xi(\tau) d \tau,\nonumber
\end{equation}
and in order to fulfill the condition of periodicity of the response, the driving has to be appropriately tuned. Since the equations of motion correspond to undamped oscillators, in general only discrete values of one cycle period $T$ are possible. Equations of motion can efficiently be solved by Fourier expansion in the time domain and then analytical solution is possible for a broad class of drivings.

\section{Circular driving}

\begin{figure}[H]
\centering
\includegraphics[width=0.3\textwidth]{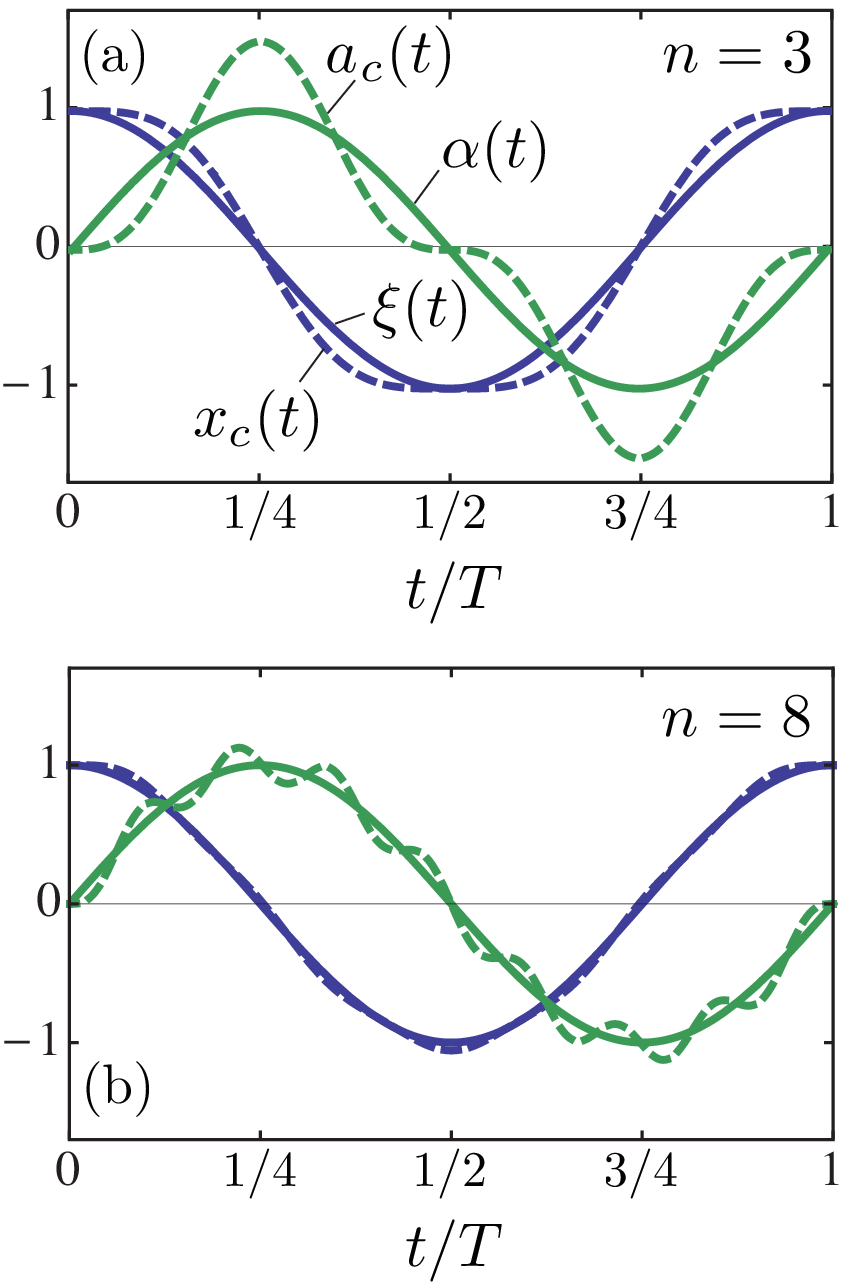}
\caption{Position $\xi(t)/\xi_0$ and the SOI $\alpha(t)/\alpha_0$ (full) as functions of driving time $t/T$ and responses $x_c(t)/\xi_0$, $a_c(t)/\alpha_0$ (dashed). Panel (a) shows fast, non-adiabatic driving, $n=3$, {\it i.e.}, $T=3T_0$, and panel (b) a slower driving with $n=8$. Fast oscillations correspond to frequency of the oscillator, $\omega=2\pi/T_0$.\label{xa}} 
\end{figure}

Here we consider driving corresponding to a circular path $\alpha[\xi]$ with constant frequency, $\xi(t) = \xi_0 \cos( \omega t/n) $ and $\alpha(t) = \alpha_0 \sin( \omega t/n)$, where $n\ge2$ is integer, the period is $T=n T_0$ and  $T_0=2\pi/\omega$.  Periodic responses with $x_c(0)=\xi(0)$ and $a_c(0)=\alpha(0)$ are given by 
\begin{eqnarray}
x_c(t)&=&\xi_0 \frac{n^2 \cos( \omega t/n)-\cos ( \omega t)}{n^2-1},\nonumber\\
a_c(t)&=&\alpha_0\frac{n \left(n \sin \left( \omega t/n\right)-\sin ( \omega t )\right)}{n^2-1}.\nonumber
\end{eqnarray}
The phases (see Eq. (10), (17) and (18)) are given analytically, 
\begin{eqnarray}
\phi_{\rm ad}&=&m^*  \!\oint_{{\cal C}_{\rm ad}} \!\! \alpha[\xi]  {d}\xi=-\pi m^*\xi_0\alpha_0,\nonumber\\
\phi_T &=&m^*  \!\oint_{{\cal C}_1}  \!\!a_c[\xi]  {d}\xi=m^*  \!\oint_{{\cal C}_2}  \!\!\alpha[x_c]  {d}x_c\nonumber\\
&=&{n^2\over n^2 - 1}  \phi_{\rm ad},\nonumber\\
\phi_c&=&m^*  \!\oint_{{\cal C}_3}  \!\!a_c[x_c]  {d}x_c=\frac{ n^2 \left(n^2+1\right) }{\left(n^2-1\right)^2}\phi_{\rm ad},\nonumber
\\
\phi_{\rm a}&=& m^* ( \oint_{{\cal C}_4} \!\!\dot{x}_c[x_c] {d}x_c + \oint_{{\cal C}_5}\!\dot{a}_c[a_c] {d}a_c/\omega^{2}  )\nonumber\\
&=&\pi m^* \frac{ n\left(n^2+1\right) \xi_0^2\omega+2 n^3\alpha_0^2/\omega}{\left(n^2-1\right)^2}.\nonumber
\end{eqnarray}
$\phi_T$ is in fact independent of the initial values $x_c(0)$, $a_c(0)$, $\dot x_c(0)$, and $\dot a_c(0)$ in spite of the fact that the contours ${\cal C}_{1,2}$ strongly depend on the choice of the initial values. Note also that $\phi_c/\phi_T>\phi_T/\phi_{\rm ad}>1$ and $\phi_{\rm a}>0$ while in the adiabatic limit, $n\to\infty$, $\phi_T=\phi_c=\phi_{\rm ad}$ and $\phi_{\rm a}=0$.  

In Fig.~\ref{xa}(a) is presented normalized displacement $\xi(t)/\xi_0$ and the SOI $\alpha(t)/\alpha_0$ as functions of driving time $t/T$ for $n=3$ (full lines). Dashed lines correspond to the response functions $x_c(t)/\xi_0$ and  $a_c(t)/\alpha_0$. The spin-orbit response exhibits a more pronounced oscillatory behavior, because the initial condition, at $t=0$, was chosen $\dot a_c(0)=0$, which differs from $\dot \alpha(0)>0$. For the displacement we chose $\dot x_c(0)=\dot \xi(0)$ which leads to a more synchronized motion. In Fig.~\ref{xa}(b) analogous results for $n=8$ are presented and here also $a_c(t)$ displays a higher degree of oscillations which would  by progressively larger $n$ diminish as ${\cal O}(1/n)$.

\begin{figure}[H]
\centering
\includegraphics[width=0.34\textwidth]{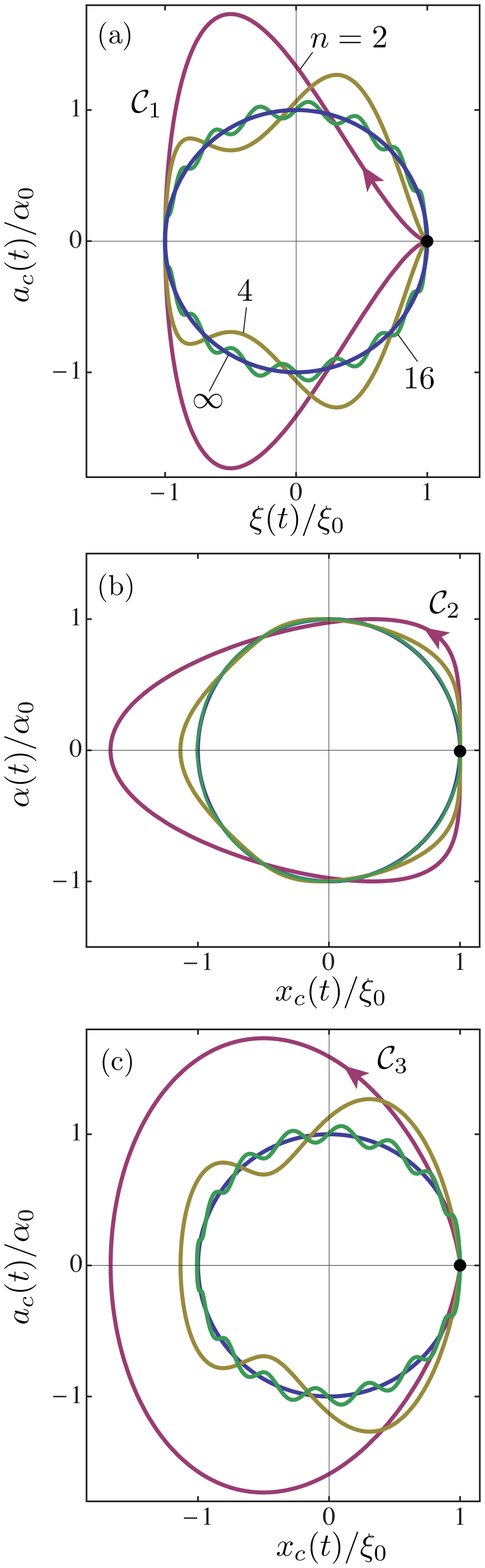}
\caption{Panel (a): Contours $[\xi(t),a_c(t)]$ or ${\cal C}_1$ (scaled) for various $n=2,4,16$ and the adiabatic limit $n\to\infty$ (circular contour ${\cal C}_{\rm ad}$). Panel (b): Contours $[x_c(t),\alpha(t)]$ corresponding to ${\cal C}_2$; colors for $n=2,4,16,\infty$ as in panel (a). Note entirely different ${\cal C}_1$ and ${\cal C}_2$, but still $\oint_{{\cal C}_1}  a_c[\xi]  {d}\xi=\oint_{{\cal C}_2}  \alpha[x_c] {d}x_c$ for each $n$. Panel (c): Contours $[x_c(t),a_c(t)]$ or ${\cal C}_3$. In all panels bullets represent start (end) of a cycle, with $x_c(0)=\xi_0$, $\dot{x}_c(0)=0$ and $a_c(0)=0$, $\dot{a}_c(0)=0$. Note also that in the adiabatic limit all contours reduce to circle ${\cal C}_{\rm ad}$.\label{fiji}}
\end{figure}

Examples of trajectories $[\xi(t),a_c(t)]$ (contours ${\cal C}_1$) are shown in Fig.~\ref{fiji}(a) from the fastest non-adiabatic $n=2$ case towards adiabatic with $n=16$ and the adiabatic limit $n\to\infty$ result corresponds to the driving $[\xi(t),\alpha(t)]$. Bullet represents the initial point at $t=0$. In Fig.~\ref{fiji}(b) the corresponding results for $[\alpha(t),x_c(t)]$ represent contours ${\cal C}_2$. Although the shapes of  ${\cal C}_1$ and ${\cal C}_2$ are different, the enclosed area for the same $n$ is equal. Note that ${\cal C}_2$ exhibits much less oscillations, consistent with  Fig.~\ref{xa}. The last of three important trajectories, $[x_c(t),a_c(t)]$, contour ${\cal C}_3$, is presented in Fig.~\ref{fiji}(c). 

\begin{figure}[H]
\centering
\includegraphics[width=0.34\textwidth]{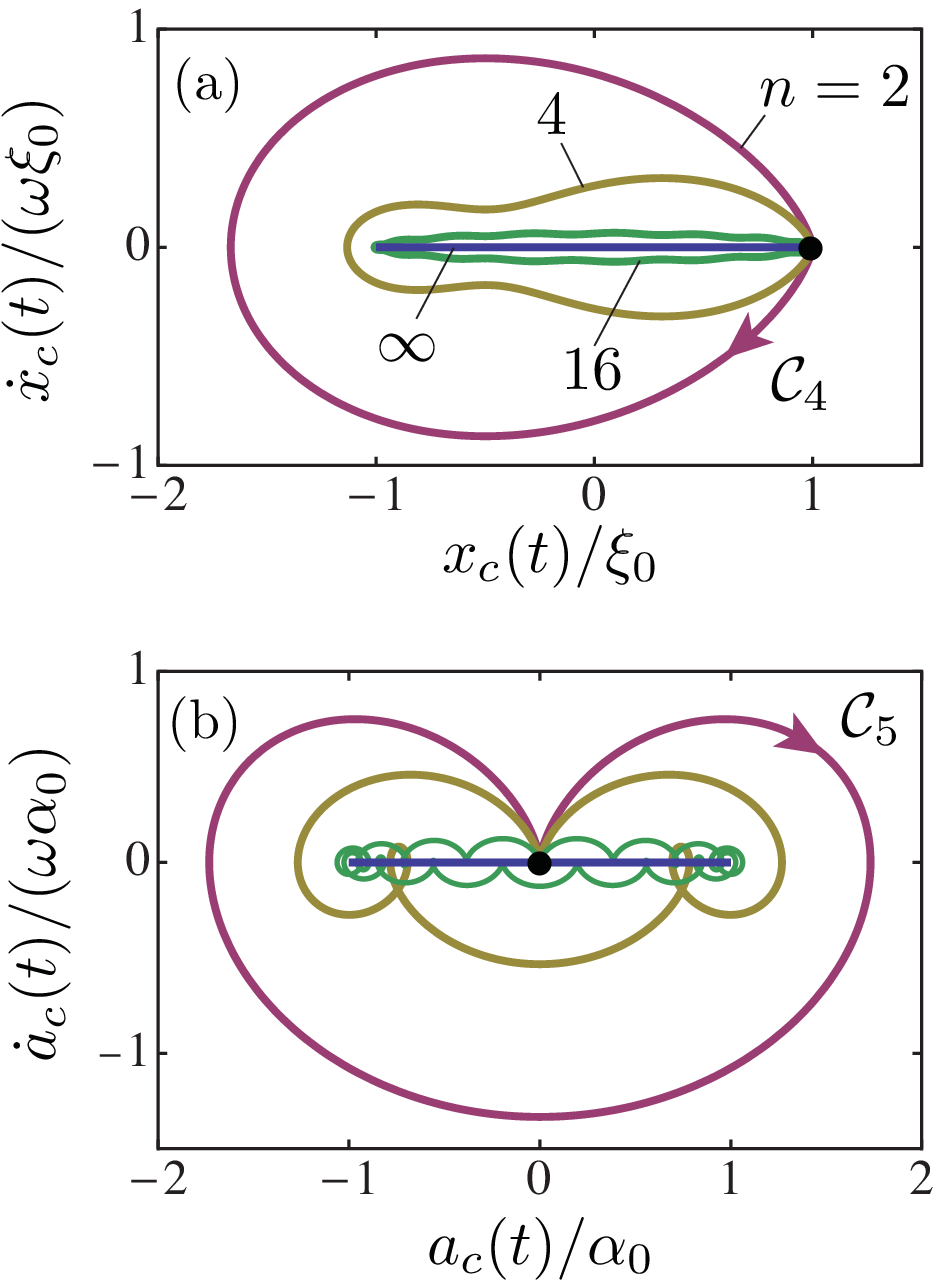}
\caption{Panel (a) shows scaled phase space contours $[x_c(t),\dot x_c(t)]$ or ${\cal C}_4$ and panel (b) $[a_c(t),\dot a_c(t)]$ or ${\cal C}_5$, for various $n=2,4,16$. In the adiabatic limit, $n\to\infty$, $\dot x_c(t)=\dot a_c(t)=0$.\label{fia}}
\end{figure}

The last set of contours corresponds to the paths in the phase space of separate, displacement and spin-orbit degrees of freedom. In Fig.~\ref{fia}(a) is shown $[x_c(t),\dot x_c(t)]$, the contour ${\cal C}_4$. In the adiabatic limit it shrinks to the line leading to vanishing enclosed area. The corresponding $[a_c(t),\dot a_c(t)]$ trajectory, the contour ${\cal C}_5$, is presented in Fig.~\ref{fia}(b), in the adiabatic limit a line decorated by superimposed oscillations.

\section{Broken ellipsoidal driving}

\begin{figure}[H]
\centering
\includegraphics[width=0.35\textwidth]{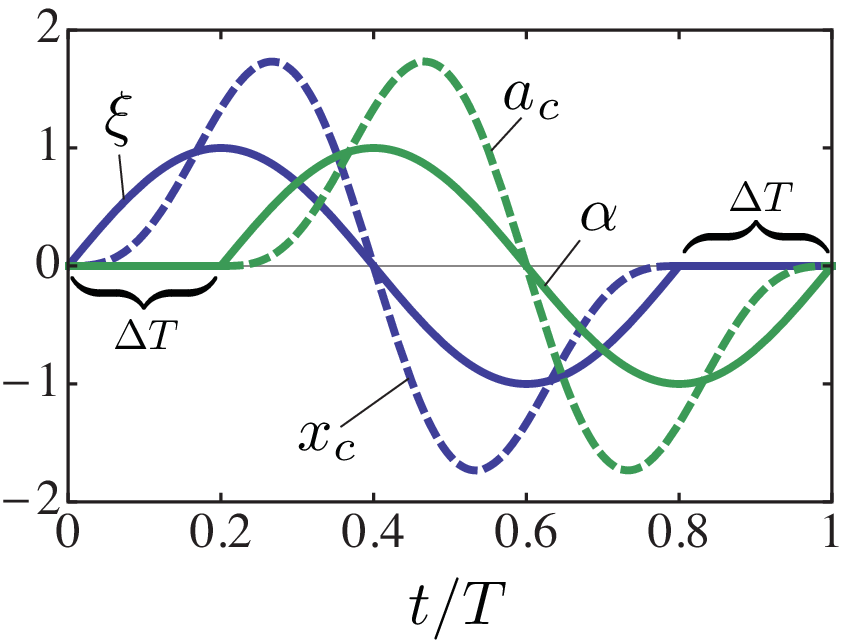}
\caption{Position $\xi(t)/\xi_0$ and the SOI $\alpha(t)/\alpha_0$ as functions of driving time $t/T$ and  the corresponding responses $x_c(t)/\xi_0$, $a_c(t)/\alpha_0$. Note that the time-dependence of SOI is delayed by $\Delta T$, otherwise with time-dependence identical to the displacement driving.\label{SH1}}
\end{figure}

Here we consider driving corresponding to  the path $\alpha[\xi]$ with $\xi(t) = \xi_0 \sin \left( \omega t/2\right) \Theta(t)\Theta(2T_0-t)$, and $\alpha(t) = {\alpha_0}\xi(t-\Delta T)/{\xi_0}$,  where $\Theta(t)$ is the Heaviside step function, $T_0=2\pi/\omega$ and $\Delta T$ is the time delay (see Fig.~\ref{SH1}). The driving is applied periodically with the cycle period $T=2T_0+\Delta T$. The responses are periodic and within one cycle are given by 
\begin{eqnarray} 
x_c(t)&=&\frac{2}{3} \xi_0\left[2 \sin \left({ \omega t }/{2}\right)-\sin (\omega t)\right]\Theta(t)\Theta(2T_0-t),\nonumber\\ 
a_c(t)&=&{\alpha_0}x_c(t-\Delta T)/{\xi_0}. \nonumber
\end{eqnarray}

\begin{figure}[H]
\centering
\includegraphics[width=0.35\textwidth]{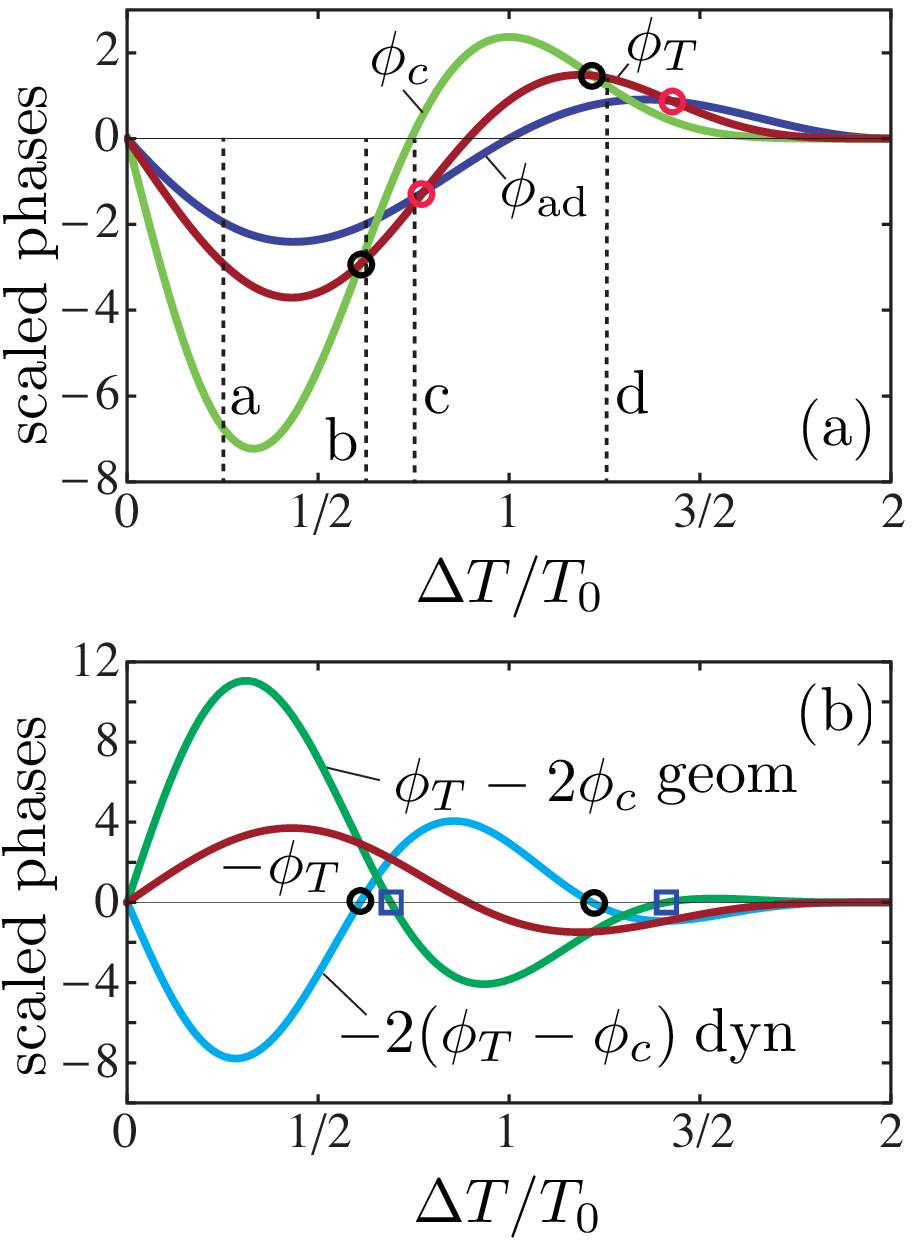}
\caption{Panel (a): All three relevant phases $\phi_T$, $\phi_c$, and $\phi_{\rm ad}$ plotted as a function of delay $\Delta T/T_0$. The phases are scaled by the factor $m^*\xi_0\alpha_0$. Black circles represent two points of equality of  $\phi_T=\phi_c$, where the dynamical phase is diagonal. Red circles indicate two $\phi_T=\phi_{\rm ad}$ points. Note that the phases change sign at different $\Delta T$. Panel (b): Off-diagonal geometric and dynamical phase, $\phi_T-2\phi_c$ or $-2(\phi_T-\phi_c)$, respectively. Note circles and squares indicating points of diagonal geometric or dynamical phase, respectively.\label{SH2}}
\end{figure}

Phases $\phi_T$, $\phi_c$, and $\phi_{\rm ad}$, calculated as a function of delay $\Delta T/T_0$, are presented in Fig.~\ref{SH2}(a). There are several interesting details to be noted: (i) all curves are similar in the sense that particular phase for small $\Delta T$ is negative and by progressively larger time delay at some point changes sign and finally vanishes at $\Delta T= 2T_0$, where there is no overlap between $\xi(t)$ and $a_c(t)$, see Fig.~\ref{SH1}. (ii) All phases mutually intersect in two $\Delta T$  points, therefore $\phi_T$ can be tuned to be equal to $\phi_c$, which eliminates off-diagonal parts of the dynamical phase, for example. It is evident also that one can tune $\phi_T=2\phi_c$ which eliminates off-diagonal parts of the geometric phase [see also Fig.~\ref{SH2}(b)]. (iii) Due to the fact that each of the phases at some point vanishes and changes sign, the ratio between any pair of phases can take any value, positive or negative. Since the amplitudes of drivings, $\xi_0$ and $\alpha_0$, are additional free parameters, consequently one can by changing $\Delta T$ tune the phases to any value -- independently.

\begin{figure}[ hbt]
\center{\includegraphics[width=0.7\textwidth]{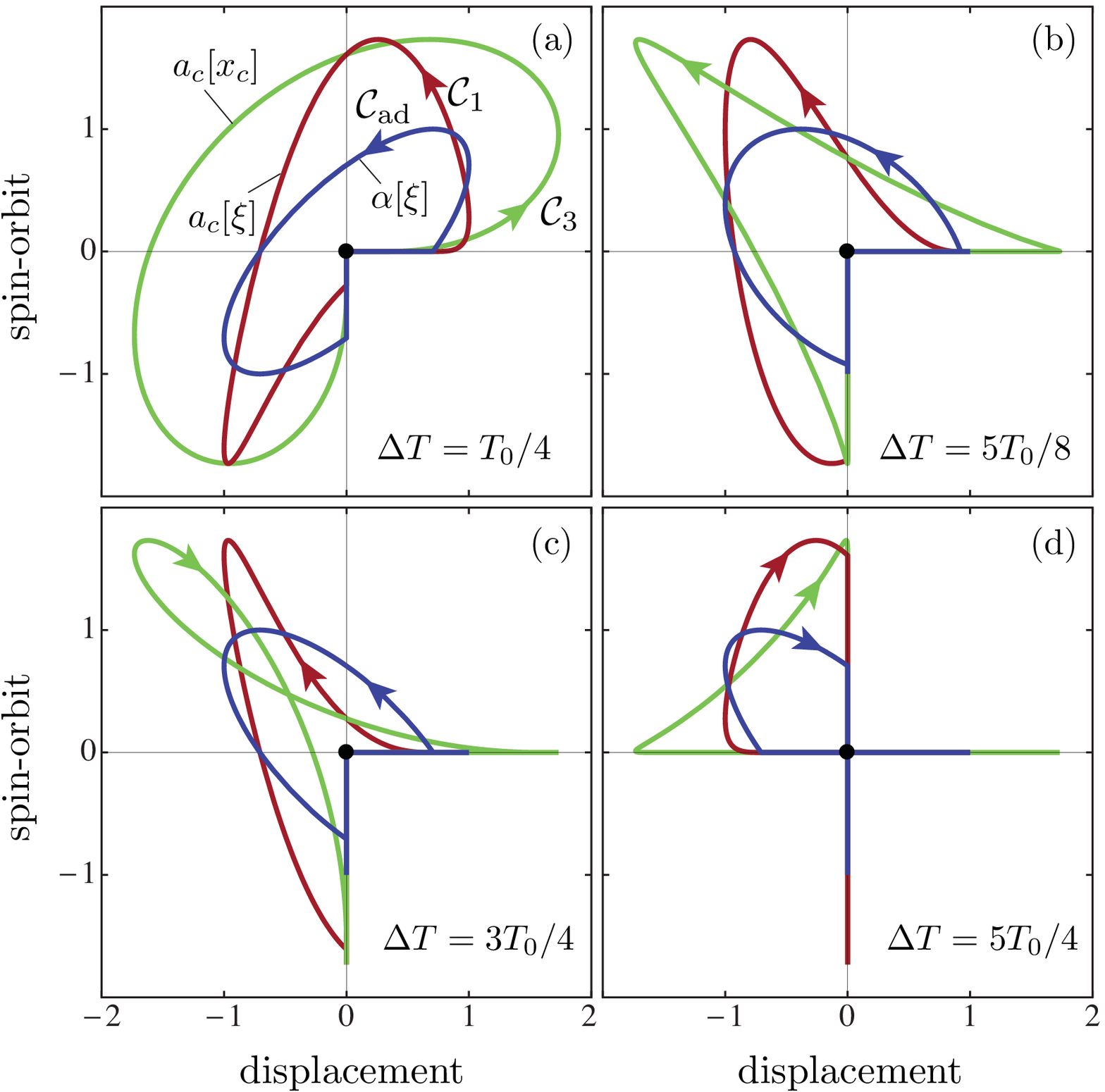}
\caption{Contours $[\xi(t)/\xi_0,a_c(t)/\alpha_0]$  (${\cal C}_1$), $[x_c(t)/\xi_0,a_c(t)/\alpha_0]$ (${\cal C}_3$) and $[\xi(t)/\xi_0,\alpha(t)/\alpha_0]$ (${\cal C}_{\rm ad}$). Panels (a), (b), (c), (d) correspond to different values of $\Delta T/T_0=1/4,5/8,3/4,5/4$, respectively (in Fig.~\ref{SH2}(a) indicated by dashed lines). \label{SH3}
}}
\end{figure}

Contours $[\xi(t)/\xi_0,a_c(t)/\alpha_0]$, $[x_c(t)/\xi_0,a_c(t)/\alpha_0]$  and $[\xi(t)/\xi_0,\alpha(t)/\alpha_0]$  are presented in Fig.~\ref{SH3}. In panels (c) and (d) note the reversion of the directions of ${\cal C}_{1,3,{\rm ad}}$ which is the reason for the change of sign of the phases shown in Fig.~\ref{SH2}. In all panels bullets represent start (end) of a cycle, with $x_c(0)=x_c(T)=0$, $\dot{x}_c(0)=\dot{x}_c(T)=0$, $a_c(0)=a_c(T)=0$, and $\dot{a}_c(0)=\dot{a}_c(T)=0$.

Scaled phase space contours $[x_c(t),\dot x_c(t)]$ and $[a_c(t),\dot a_c(t)]$ are identical to the case of circular motion for SOI response $[a_c(t),\dot a_c(t)]$, Fig.~\ref{fia}(b) for $n=2$, with the displacement response appropriately scaled by $\xi_0/\alpha_0$.

\break\widetext
~~~~~~~~~~~~~~~~~~~~~~~~~~~~~~~~~~~~~~~~~~~~~~~~~~~~~~~~~~~~~~~~~~~~~~~~~~~~~~~~~~~~~~~~~~~~~~~~~~~~~~~~~~~~~~~~~~~~~~~~~~~~~~~~~~~

\end{document}